\documentclass[12pt,a4paper]{article}
\usepackage{amsmath}
\usepackage{amsfonts}
\usepackage{cite}

\renewcommand{\thesection}
  {\arabic{section}.\hspace{-.5em}}

\makeatletter
\renewcommand\section{
  \@startsection{section}{3}{\z@}%
  {-3.25ex\@plus -1ex \@minus -.2ex}%
  {1.5ex \@plus .2ex}%
  {\normalfont\normalsize\bfseries\mathversion{bold}}}
\renewcommand\subsection{
  \@startsection{subsection}{3}{\z@}%
  {-3.25ex\@plus -1ex \@minus -.2ex}%
  {1.5ex \@plus .2ex}%
  {\normalfont\normalsize\bfseries\mathversion{bold}}}
\makeatother

\makeatletter \@addtoreset{equation}{section} \makeatother
\renewcommand{\theequation}{\arabic{section}.\arabic{equation}}

\makeatletter
\renewcommand{\appendix}{
\renewcommand{\thesection}{\Alph{section}.\hspace{-.5em}}
\@addtoreset{equation}{section}
\renewcommand{\theequation}{\Alph{section}.\arabic{equation}}
\setcounter{section}{0}}
\makeatother

\newcommand{\Eqn}[1]{&\hspace{-0.5em}#1\hspace{-0.5em}&}
\newcommand{\nn}{\nonumber}
\renewcommand{\[}{\begin{equation}}
\renewcommand{\]}{\end{equation}}
\newcommand{\eqb}{\begin{eqnarray}}
\newcommand{\eqe}{\end{eqnarray}}

\newcommand{\bbZ}{{\mathbb Z}}
\newcommand{\bbP}{{\mathbb P}}

\newcommand{\grp}[1]{\mathrm{#1}}
\newcommand{\varth}{\vartheta}

\newcommand{\Rt}{R^\vee}
\newcommand{\bfm}{\boldsymbol{m}}
\newcommand{\bfv}{\boldsymbol{v}}
\newcommand{\bfw}{\boldsymbol{w}}
\newcommand{\bfmu}{\boldsymbol{\mu}}
\newcommand{\bfR}{\boldsymbol{R}}
\newcommand{\bfzero}{\boldsymbol{0}}
\newcommand{\ta}{\tilde{a}}
\newcommand{\tb}{\tilde{b}}
\newcommand{\tj}{\tilde{\jmath}}
\newcommand{\tu}{\tilde{u}}
\newcommand{\tx}{\tilde{x}}
\newcommand{\ty}{\tilde{y}}
\newcommand{\ttau}{\tilde{\tau}}
\newcommand{\calZ}{{\cal Z}}
\newcommand{\ZSQCD}[2]{\calZ^{\grp{SU}(#1)}_{N_{\rm f}=#2}}
\newcommand{\gA}[1]{\alpha_{#1}}
\newcommand{\gB}[1]{\beta_{#1}}
\newcommand{\gL}[1]{\lambda_{#1}}
\newcommand{\tgL}[1]{\tilde{\lambda}_{#1}}

\begin{document}


\def\papertitlepage{\baselineskip 3.5ex \thispagestyle{empty}}
\def\preprinumber#1#2{\hfill
\begin{minipage}{1.0in}
#1 \par\noindent #2
\end{minipage}}

%
\papertitlepage
\setcounter{page}{0}
\preprinumber{YITP-12-59}{} 
\vskip 2ex
\vfill
\begin{center}
{\large\bf\mathversion{bold}
Seiberg--Witten prepotential for E-string theory\\[.5ex]
and global symmetries
}
\end{center}
\vfill
\baselineskip=3.5ex
\begin{center}
  Kazuhiro Sakai\\

{\small
\vskip 6ex
{\it Yukawa Institute for Theoretical Physics, Kyoto University}\\[1ex]
{\it Kyoto 606-8502, Japan}\\
\vskip 1ex
{\tt ksakai@yukawa.kyoto-u.ac.jp}

}
\end{center}
\vfill
\baselineskip=3.5ex
\begin{center} {\bf Abstract} \end{center}

We obtain Nekrasov-type expressions
for the Seiberg--Witten prepotential for
the six-dimensional (1,0) supersymmetric
E-string theory compactified on $T^2$
with nontrivial Wilson lines.
We consider compactification with four general Wilson line parameters,
which partially 
break the $E_8$ global symmetry.
In particular, we investigate in detail
the cases where the Lie algebra of the unbroken global symmetry is
$E_n\oplus A_{8-n}$ with $n=8,7,6,5$ or $D_8$.
All our Nekrasov-type expressions
can be viewed as special cases of
the elliptic analogue of the Nekrasov partition function
for the $\grp{SU}(N)$ gauge theory with $N_{\rm f}=2N$ flavors.
We also present a new expression for
the Seiberg--Witten curve for the E-string theory
with four Wilson line parameters,
clarifying the connection between the E-string theory and
the $\grp{SU}(2)$ Seiberg--Witten theory with
$N_{\rm f}=4$ flavors.

\vfill
\noindent
July 2012


\setcounter{page}{0}
\newpage
\renewcommand{\thefootnote}{\arabic{footnote}}
\setcounter{footnote}{0}
\setcounter{section}{0}
\baselineskip = 3.5ex
\pagestyle{plain}
%

\section{Introduction}

The E-string theory is one of the simplest interacting
quantum field theories with (1,0) supersymmetry in six dimensions
\cite{Ganor:1996mu,Seiberg:1996vs,Klemm:1996hh,
      Ganor:1996pc,Minahan:1998vr}.
It is obtained as the low energy theory of the heterotic string theory
on K3 when an $E_8$ instanton shrinks to zero size
\cite{Ganor:1996mu,Seiberg:1996vs}.
The theory is decoupled from gravity.
It is probably a conventional local quantum field theory,
though it does not have a Lagrangian description.
Another unusual feature is that
fundamental excitations are strings, called E-strings,
rather than particles.
The moduli space of vacua consists of
a Coulomb branch with one tensor multiplet
and a Higgs branch with 29 hypermultiplets.
There are no vector multiplets and
$E_8$ appears as a global symmetry group.

The E-string theory shows extremely rich properties
when toroidally compactified down to lower dimensions
\cite{Klemm:1996hh,Ganor:1996pc,Minahan:1998vr,Ganor:1996xd,
      Morrison:1996xf,Douglas:1996xp,Lerche:1996ni,Minahan:1997ch,
      Minahan:1997ct, Mohri:2001zz}.
In the compactified theories
one can break the $E_8$ global symmetry
by coupling its currents to the background $E_8$ gauge field
with nontrivial Wilson lines.
For each circle of the toroidal compactification
there are eight Wilson line parameters
taking their values in the Cartan torus of $E_8$.
By turning on these parameters, one can break $E_8$
to its subgroups and realize models with various global symmetries.

When the theory is toroidally compactified down to four dimensions,
the low energy dynamics in the Coulomb branch
is described by Seiberg--Witten theory
\cite{Seiberg:1994rs,Seiberg:1994aj}.
The Seiberg--Witten curve was constructed
with the most general Wilson line parameters
\cite{Ganor:1996pc,Eguchi:2002fc}.
Recently, it was found that the Seiberg--Witten prepotential
admits a Nekrasov-type expression \cite{Sakai:2012zq}.
The expression is for the case with no Wilson line parameters.
In this paper we present a Nekrasov-type expression
with four general Wilson line parameters.
It is verified up to a sufficiently high order
(involving Young diagrams with 10 boxes) that
the prepotential given by this expression is
in perfect agreement with
that computed from the Seiberg--Witten curve.

The Seiberg--Witten curve with full eight Wilson line parameters
takes a rather complicated form. 
In the case of four Wilson line parameters,
however, the curve reduces to a much simpler expression.
Interestingly, it is expressed in terms of the curve for
the $\grp{SU}(2)$ Seiberg--Witten theory with
$N_{\rm f}=4$ flavors.
It has been known that the low energy theory of
the E-string theory on $T^2$ can flow to that of the
four-dimensional $\grp{SU}(2)\ N_{\rm f}=4$ theory.
In particular, it was argued that
the $\grp{SL}(2,\bbZ)$ duality of the latter theory
is derived from
the $\grp{SL}(2,\bbZ)$ action
on the $T^2$ \cite{Ganor:1996pc}.
Our expression clarifies
how it occurs, in particular how the $\grp{SO}(8)$ triality
emerges from the E-string theory.

By adjusting four Wilson line parameters to special values,
one can realize the cases where
the Lie algebra of the unbroken global symmetry is
$E_n\oplus A_{8-n}$ with $n=8,7,6,5$ or $D_8$.
We present explicit forms of the Seiberg--Witten curves
and the Nekrasov-type expressions for these specific cases.
In each of these cases the Seiberg--Witten prepotential
counts multiplicities of BPS E-strings wound around
one of the circles of the toroidal compactification
with general winding numbers and momenta.
The multiplicities are equivalent to
Gromov--Witten invariants associated with
the $E_n$ del Pezzo surface or $\bbP^1\times \bbP^1$
embedded in a Calabi--Yau threefold
\cite{Klemm:1996hh,Lerche:1996ni,Chiang:1999tz,Katz:1999xq}.
Our Nekrasov-type expressions
provide us with the generating functions 
of these invariants in very simple, closed forms.
In particular, these expressions
respect the modular properties of
the partition functions of wound BPS E-strings
\cite{Minahan:1997ct,Minahan:1998vr,Mohri:2001zz}.

As in the case with no Wilson line parameters \cite{Sakai:2012zq},
our Nekrasov-type expression can be viewed as a special case
of the elliptic analogue of
the Nekrasov partition function for the
$\grp{SU}(4)$ gauge theory with $N_{\rm f}=8$ flavors
\cite{Nekrasov:2002qd,Hollowood:2003cv}.
Moreover, for particular
values of the Wilson line parameters
our expression can be embedded in
the elliptic analogue of the Nekrasov partition functions
for the $\grp{SU}(N)\ N_{\rm f}=2N$ theories
with $N=3,2$.
In fact, for all the cases with
$E_n\oplus A_{8-n}$ with $n=8,7,6,5$ and $D_8$
the Nekrasov-type expressions can be viewed as special cases of
the elliptic Nekrasov partition function
for the $\grp{SU}(3)\ N_{\rm f}=6$ theory.
Furthermore, expressions for the cases with
$E_7\oplus A_1$,
$E_5\oplus A_3$
and $D_8$ can also be viewed as special cases
of the elliptic Nekrasov partition function
for the $\grp{SU}(2)\ N_{\rm f}=4$ theory.

The organization of this paper is as follows.
In section~2, we present our new expression for
the Seiberg--Witten curve with four general Wilson line parameters
and discuss its properties.
In section~3, we present the Nekrasov-type expression
with four Wilson line parameters.
We then focus on some particular cases in which
the general formula reduces to a sum over fewer partitions.
In section~4,
we investigate in detail the cases with global symmetries
$E_n\oplus A_{8-n}$ with $n=8,7,6,5$ and $D_8$.
Conventions of special functions
are summarized in Appendix~A.

\section{Seiberg--Witten curve with four Wilson line parameters}

In this section we present a new expression for the Seiberg--Witten
curve for the E-string theory compactified on $T^2$ with
four Wilson line parameters.
We clarify how it is related to the
Seiberg--Witten curve for the four-dimensional
$\grp{SU}(2)$ gauge theory with $N_{\rm f}=4$ flavors.

The Seiberg--Witten curve for the E-string theory
compactified on $T^2$
with the most general Wilson line parameters
was constructed in \cite{Ganor:1996pc,Eguchi:2002fc}.
(An improved expression 
in terms of $E_8$-invariant Jacobi forms
is available in \cite{Sakai:2011xg}.)
It takes the following form
\[\label{geneE8curve}
y^2=4x^3-fx-g
\]
with
\[\label{genefg}
f=\sum_{j=0}^4 a_j u^{4-j},\qquad
g=\sum_{j=0}^6 b_j u^{6-j}.
\]
The coefficients $a_j,b_j$ depend on 
nine complex parameters,
$\tau$ and $\bfmu=(\mu_1,\ldots,\mu_8)$.
$\tau$ is the complex modulus of the $T^2$
and $\bfmu$ are the Wilson line parameters
which specify the background $E_8$ gauge field
along the $T^2$ \cite{Ganor:1996xd}.
In this paper we restrict ourselves to the cases
in which Wilson line parameters take the form
\[
\label{WLdoubleD4}
\bfmu=(m_1,m_2,m_3,m_4,m_1,m_2,m_3,m_4)
\]
or
\[
\label{WLmlD4}
\bfmu=(0,0,0,0,m_1+m_2,m_1-m_2,m_3+m_4,m_3-m_4).
\]
These two configurations are related to each other
by a sequence of $E_8$ Weyl reflections
and thus correspond to the same curve.
Keeping this embedding in mind,
we hereafter specify the Wilson line parameters
by the following short notation
\[
\bfm=(m_1,m_2,m_3,m_4).
\]

The most general Seiberg--Witten curve for the E-string theory
is invariant under the automorphism group consisting of
affine $E_8$ Weyl group
and $\grp{SL}(2,\bbZ)$ \cite{Eguchi:2002nx}.
From this one can easily deduce the automorphism
of the curve in the present setup as follows.
The curve with four Wilson line parameters is invariant
under the following transformations
\eqb
\label{minterchange}
\Eqn{\bullet} m_i\leftrightarrow m_j\qquad\mbox{for any}\ i\ne j,\\
\Eqn{\bullet} m_i\leftrightarrow -m_i \qquad\mbox{for any}\ i,\\
\Eqn{\bullet} m_i\rightarrow m_i-\frac{1}{2}\sum_{j=1}^4 m_j
  \qquad\mbox{for all}\ i=1,\ldots,4,\\
\Eqn{\bullet} \bfm\rightarrow \bfm+\bfw,
  \qquad\bfw\in \Gamma_{\rm w}.
\label{mtransl}
\eqe
Here $\Gamma_{\rm w}$ denotes the weight lattice of $D_4$,
\[
 \Gamma_{\rm w}
  :=\bigl\{\bfw
  \in\bbZ^4\cup\bigl(\bbZ+\tfrac{1}{2}\bigr)^4\bigr\}.\\
\]
The group generated by the above transformations is in fact
the affine automorphism group of $\Gamma_{\rm w}$.
There is also an $\grp{SL}(2,\bbZ)$ automorphism
generated by the following transformations
\eqb
\Eqn{\bullet}\tau\to\tau+1,\\
\Eqn{\bullet}\tau\to -\frac{1}{\tau},\ \bfm\to\frac{\bfm}{\tau}
  \quad\mbox{with}\quad(u,x,y)\to
  (\tau^{-6}Lu,\tau^{-10}L^2x,\tau^{-15}L^3y),
  \quad
\eqe
where
\[
L:=e^{2\pi i|\bfm|^2/\tau}.
\]

In principle,
the explicit form of the curve with four Wilson line parameters
is obtained by simply substituting
(\ref{WLdoubleD4}) or (\ref{WLmlD4}) into the expression in
\cite{Sakai:2011xg}.
However, the expression constructed in this way
is rather complicated for practical purposes.
In what follows we will express the same Seiberg--Witten curve
in a more convenient form by means of the curve
for the $\grp{SU}(2)$ gauge theory
with $N_{\rm f}=4$ fundamental hypermultiplets.
Recall that the Seiberg--Witten
curve for the $\grp{SU}(2)\ N_{\rm f}=4$ theory 
is given by \cite{Seiberg:1994aj}
\[
\label{N_f=4curve}
\ty^2=W_1W_2W_3+A(W_1T_1(e_2-e_3)+W_2T_2(e_3-e_1)+W_3T_3(e_1-e_2))-A^2N
\]
with
\eqb
W_i\Eqn{=}\tx-e_i \tu-e_i^2 R,\nn\\
A\Eqn{=}(e_1-e_2)(e_2-e_3)(e_3-e_1),\nn\\
R\Eqn{=}\frac{1}{2}\sum_i M_i^2,\nn\\
T_1\Eqn{=}\frac{1}{12}\sum_{i>j}M_i^2 M_j^2
  -\frac{1}{24}\sum_i{M_i}^4,\nn\\
T_2\Eqn{=}-\frac{1}{2}\prod_iM_i
  -\frac{1}{24}\sum_{i>j}M_i^2 M_j^2+\frac{1}{48}\sum_i M_i^4,\nn\\
T_3\Eqn{=} \frac{1}{2}\prod_iM_i
  -\frac{1}{24}\sum_{i>j}M_i^2 M_j^2+\frac{1}{48}\sum_iM_i^4,\nn\\
N\Eqn{=}\frac{3}{16}\sum_{i>j>k}M_i^2 M_j^2 M_k^2
  -\frac{1}{96}\sum_{i\ne j}M_i^2 M_j^4+\frac{1}{96}\sum_i M_i^6,\nn\\
\label{e1e2e3}
e_1\Eqn{=}\frac{\varth_3^4+\varth_4^4}{12},\qquad
e_2=\frac{\varth_2^4-\varth_4^4}{12},\qquad
e_3=\frac{-\varth_2^4-\varth_3^4}{12}.
\eqe
Here $\varth_k:=\varth_k(0,\tau)$ are the Jacobi theta functions
(see Appendix A).
$\tau$ denotes the complexified bare gauge coupling
and $M_1,\ldots,M_4$ are the masses of
the fundamental hypermultiplets.
To obtain the Seiberg--Witten curve for the E-string theory,
let us first 
make the following transformation of variables,
\eqb
\label{formaltransf}
\tu\Eqn{=}-\frac{\eta^{24}}{l^2}(u+u_0)-\frac{E_6}{12E_4}R,\nn\\
\tx\Eqn{=}-\frac{\eta^{24}}{l^2(u-u_0)}x+\frac{E_4}{72}R,\nn\\
\ty^2\Eqn{=}-\frac{\eta^{72}}{4l^6(u-u_0)^3}y^2.
\eqe
Here $E_{2k}:=E_{2k}(\tau)$ and $\eta:=\eta(\tau)$
are the Eisenstein functions and the Dedekind eta function
respectively.
The curve (\ref{N_f=4curve})--(\ref{e1e2e3})
can then be written in the form
\[\label{6dD4curve}
y^2=4x^3-\left(\ta_0u^2+\ta_1u+\ta_2\right)(u-u_0)^2 x 
-\left(\tb_0u^3+\tb_1u^2+\tb_2u+\tb_3\right)(u-u_0)^3.
\]
Here
$\ta_j,\tb_j$ are some functions in
$\tau$, $l M_i$ and $u_0$.
$l$ is a parameter that gives an inverse mass scale.
It can be absorbed in the definitions
of $\tu,\tx,\ty$ and $M_i$, but
let us keep it for later use.
Next, we identify the parameters as
\eqb\label{u0def}
u_0
\Eqn{=}
\frac{1}{2\eta^{12}E_4}\sum_{\sigma\in {\rm S}_4}
\prod_{j=1}^4\varth_j(m_{\sigma(j)},\tau)^2
\eqe
and
\eqb
\label{Mthetarel}
l M_1
   \Eqn{=}\prod_{j=1}^4\varth_1(m_j,\tau)
         -\prod_{j=1}^4\varth_2(m_j,\tau),\nn\\
l M_2
   \Eqn{=}\prod_{j=1}^4\varth_1(m_j,\tau)
         +\prod_{j=1}^4\varth_2(m_j,\tau),\nn\\
l M_3
   \Eqn{=}\prod_{j=1}^4\varth_3(m_j,\tau)
         -\prod_{j=1}^4\varth_4(m_j,\tau),\nn\\
l M_4
   \Eqn{=}\prod_{j=1}^4\varth_3(m_j,\tau)
         +\prod_{j=1}^4\varth_4(m_j,\tau).
\eqe
In (\ref{u0def}),
$\sigma$ denotes a permutation of $\{1,2,3,4\}$ and 
the sum is taken over all such permutations.
Under this identification
the curve (\ref{6dD4curve})
coincides precisely with the Seiberg--Witten curve for
E-string theory \cite{Eguchi:2002fc,Sakai:2011xg}
with the Wilson line parameters
given by (\ref{WLdoubleD4}) or (\ref{WLmlD4}).

By reversing the above construction,
one can reproduce the Seiberg--Witten curve for
the $\grp{SU}(2)\ N_{\rm f}=4$ theory
from that of the E-string theory.
The reader might think that
the transformation (\ref{formaltransf}) is artificial
because $x$ and $y$ are rescaled by $u$-dependent factors.
One could use the following linear transformation
\eqb
\label{rescalingtransf}
u\Eqn{=}-\frac{l^2}{\eta^{24}}
  \left(\tu+\frac{E_6}{12E_4}R\right)-u_0,\nn\\
x\Eqn{=}\frac{2l^2u_0}{\eta^{24}}
  \left(\tx-\frac{E_4}{72}R\right),\nn\\
y^2\Eqn{=}\frac{32l^6u_0^3}{\eta^{72}}\ty^2,
\eqe
instead of (\ref{formaltransf}).
The curve (\ref{N_f=4curve}) is then obtained
by taking the limit $l\to 0$.
In fact, under this limit the identification (\ref{formaltransf})
coincides with (\ref{rescalingtransf}).

In the above reduction the bare gauge coupling $\tau$
of the $\grp{SU}(2)\ N_{\rm f}=4$ theory
is identified with the complex modulus $\tau$ of $T^2$
on which the E-string theory is compactified.
This means that the $\grp{SL}(2,\bbZ)$ duality of
the $N_{\rm f}=4$ theory is identified with
the $\grp{SL}(2,\bbZ)$ action of the $T^2$
\cite{Ganor:1996pc}.
Recall that the $\grp{SL}(2,\bbZ)$ duality of
the $\grp{SU}(2)\ N_{\rm f}=4$ theory
is mixed with $\grp{SO}(8)$ triality \cite{Seiberg:1994aj}.
That is, the spectrum of the theory
is not invariant under
the transformations $\tau\to\tau+1$ and $\tau\to -1/\tau$,
but is invariant under the combinations
\renewcommand{\arraystretch}{1.5}
\begin{align}
&\bullet\ \tau\to\tau+1&&\mbox{with}&&
\begin{array}{l}
M_1\to M_1,\\
M_2\to M_2,\\
M_3\to M_3,\\
M_4\to -M_4
\end{array}\\
\intertext{and}
&\bullet\ \tau\to-\frac{1}{\tau}&&\mbox{with}&&
\begin{array}{l}
M_1\to \frac{1}{2}(M_1+M_2+M_3-M_4),\\
M_2\to \frac{1}{2}(M_1+M_2-M_3+M_4),\\
M_3\to \frac{1}{2}(M_1-M_2+M_3+M_4),\\
M_4\to \frac{1}{2}(-M_1+M_2+M_3+M_4).
\end{array}
\end{align}
\renewcommand{\arraystretch}{1}%
In (\ref{Mthetarel})
$M_i$ are identified with functions in $\tau$ and $m_j$.
Modular transformations of these functions
precisely reproduce the above transformations of $M_i$
(up to an overall factor which can be absorbed into $l$).
This peculiar identification
was first found in \cite{Eguchi:2002nx},
where a different connection
between the two theories was considered.

The identification (\ref{Mthetarel}) admits
the following interpretation in connection with
the automorphism group of the curve.
We saw that the automorphism group
of the present Seiberg--Witten curve is governed by
the weight lattice of $D_4$ denoted by $\Gamma_{\rm w}$.
This lattice can be viewed as the union of four sublattices
\[
\Gamma_{\rm w}
=\Gamma_{\rm b}\cup\Gamma_{\rm v}\cup\Gamma_{\rm s}\cup\Gamma_{\rm c},
\]
where
\eqb
\Gamma_{\rm b}
  \Eqn{:=}\bigl\{\bfw=(w_1,w_2,w_3,w_4)\in\bbZ^4
  \bigm| \textstyle\sum_{j=1}^4 w_j\in 2\bbZ\bigr\},\nn\\
\Gamma_{\rm v}
  \Eqn{:=}\bigl\{\bfw=
  (1,0,0,0)+\bfv
   \bigm| \bfv\in\Gamma_{\rm b}\bigr\},\nn\\
\Gamma_{\rm s}
  \Eqn{:=}\bigl\{\bfw=
  \bigl(\tfrac{1}{2},\tfrac{1}{2},\tfrac{1}{2},\tfrac{1}{2}\bigr)+\bfv
   \bigm| \bfv\in\Gamma_{\rm b}\bigr\},\nn\\
\Gamma_{\rm c}
  \Eqn{:=}\bigl\{\bfw=
  \bigl(-\tfrac{1}{2},\tfrac{1}{2},\tfrac{1}{2},\tfrac{1}{2}\bigr)+\bfv
   \bigm| \bfv\in\Gamma_{\rm b}\bigr\}.
\eqe
In terms of these sublattices,
(\ref{Mthetarel}) can be expressed as
\begin{align}
\hspace{3em}
l M_1&=-2\Theta_{\rm c}(\tau,\bfm),&
l M_2&=2\Theta_{\rm s}(\tau,\bfm),\hspace{5em}\nn\\
l M_3&=2\Theta_{\rm v}(\tau,\bfm),&
l M_4&=2\Theta_{\rm b}(\tau,\bfm),
\label{MinTheta}
\end{align}
where $\Theta_{\cal R}(\tau,\bfm)$
is the theta function for sublattice $\Gamma_{\cal R}$,
\[
\Theta_{\cal R}(\tau,\bfm):=\sum_{\bfw\in\Gamma_{\cal R}}
\exp\left(\pi i\bfw^2\tau+2\pi i\bfw\cdot\bfm\right).
\]
Note that $\Theta_{\cal R}(\tau,\bfm)/\eta(\tau)^4$
with ${\cal R}={\rm b,v,s,c}$ respectively give the characters
of the basic, vector, spinor, conjugate-spinor representations
of the affine $D_4$ algebra.
Therefore (\ref{MinTheta}) means that
the masses of the hypermultiplets
in $\grp{SU}(2)\ N_{\rm f}=4$ theory
are essentially identified with these affine $D_4$ characters.

Note that the $D_4$ symmetry acting on
the Wilson line parameters $m_j$
should not be confused with the $D_4$ symmetry
acting on the masses $M_i$.
The two $D_4$ symmetries are related in a nontrivial manner.
For instance, the exchange of $M_2$ for $M_3$ 
is an element of the Weyl group of the latter $D_4$.
We see from (\ref{MinTheta}) that this corresponds to the
exchange of $\Gamma_{\rm s}$ for $\Gamma_{\rm v}$,
which is an outer automorphism of the former $D_4$.

In the rest of this section let us sketch out how
to calculate the prepotential from the Seiberg--Witten curve.
Our Seiberg--Witten curve given by
(\ref{N_f=4curve})--(\ref{Mthetarel})
is expressed in the Weierstrass form.
An elliptic curve in the Weierstrass form
can be parametrized as
\[
y^2=4x^3-
\frac{1}{12}\frac{E_4(\ttau)}{\omega^4}x
-\frac{1}{216}\frac{E_6(\ttau)}{\omega^6}.
\]
Here $\ttau$ is the complex structure modulus
and $\omega$ (multiplied by $2\pi$)
is one of the fundamental periods
of the elliptic curve.
By comparing this expression with the explicit form of
the Seiberg--Witten curve,
one can calculate $\omega(u,\tau,\bfm),\ttau(u,\tau,\bfm)$
as series expansions in $1/u$.
They are related to the scalar vev $\varphi$
and the prepotential $F_0$ by
\eqb
\partial_u\varphi \Eqn{=} \frac{i}{2\pi}\omega,\\
\partial_\varphi^2 F_0 \Eqn{=}8\pi^3 i(\ttau-\tau).
\eqe
These relations parametrically determine the function
$F_0(\varphi,\tau,\bfm)$.
The integration constants
are determined accordingly.
The reader is referred for the details of these calculations
to \cite{Sakai:2011xg}.

\section{Nekrasov-type expression with four Wilson line parameters}

In this section we present an explicit expression
for the Seiberg--Witten prepotential for the E-string
theory with four Wilson lines and discuss its properties.

\subsection{General expression}

Let $\bfR^{(N)}=(R_1,\ldots,R_N)$ denote an $N$-tuple of
partitions. Each partition $R_k$ is a nonincreasing sequence
of nonnegative integers
\[
R_k =\{
\nu_{k,1}\ge\nu_{k,2}\ge\cdots\ge\nu_{k,\ell(R_k)}>
\nu_{k,\ell(R_k)+1}=\nu_{k,\ell(R_k)+2}=\cdots=0\}.
\]
Here the number of nonzero $\nu_{k,i}$ is
denoted by $\ell(R_k)$. $R_k$ is represented by a Young diagram.
We let $|R_k|$ denote the size of $R_k$, i.e.~the number of boxes
in the Young diagram of $R_k$:
\[
|R_k| := \sum_{i=1}^\infty\nu_{k,i}=\sum_{i=1}^{\ell(R_k)}\nu_{k,i}.
\]
Similarly, the size of $\bfR^{(N)}$ is denoted by
\[
|\bfR^{(N)}| := \sum_{k=1}^N|R_k|.
\]
We let
$\Rt_k=\{\nu_{k,1}^\vee\ge\nu_{k,2}^\vee\ge\cdots\}$ denote
the conjugate partition of $R_k$. We also introduce the notation
\[
h_{k,l}(i,j) := \nu_{k,i}+\nu_{l,j}^\vee-i-j+1,
\]
which represents the relative hook-length of a box
at $(i,j)$ between the Young diagrams of $R_k$ and $R_l$.

In our expression we consider a sum over four partitions.
For our present purpose, it is convenient to express these
partitions as
\[
\bfR^{(4)} = (R_1,R_2,R_3,R_4) = (R_{11},R_{10},R_{00},R_{01}).
\]
The prepotential is then given by
\[\label{F0Estring}
F_0 = (2\hbar^2\ln \calZ)\big|_{\hbar=0}\,,
\]
where
\[
\label{ZEstring}
\calZ=
\sum_{\bfR^{(4)}}
Q^{|\bfR^{(4)}|}
\prod_{a,b,c,d}\,
\prod_{(i,j)\in R_{ab}}
\frac
{\varth_{ab}
 \left(\tfrac{1}{2\pi}(j-i)\hbar+m_{cd},\tau\right)
 \varth_{ab}
 \left(\tfrac{1}{2\pi}(j-i)\hbar-m_{cd},\tau\right)
}
{\varth_{1-|a-c|,1-|b-d|}
 \left(\tfrac{1}{2\pi}h_{ab,cd}(i,j)
       \hbar,\tau\right)^2}\qquad
\]
and
\[
Q:=e^{2\pi i\varphi+\pi i\tau}.
\]
Here the sum is taken over all possible
partitions $\bfR^{(4)}$ (including the empty partition).
Indices $a,b,c,d$ take values $0,1$, while a set of indices
$(i,j)$ run over the coordinates of all boxes
in the Young diagram of $R_{ab}$.
$\varth_{ab}(z,\tau)$ are the Jacobi theta functions
(see Appendix A).
$h_{ab,cd}(i,j)$ is the relative hook-length
defined between partitions $R_{ab}$ and $R_{cd}$.
$m_{ab}$ are the Wilson line parameters, which are identified
with those appearing in the Seiberg--Witten curve by
\[
\bfm=(m_1,m_2,m_3,m_4)=(m_{11},m_{10},m_{00},m_{01}).
\]
If we set $\bfm=\bfzero$, 
the expression reduces to the one
studied in \cite{Sakai:2012zq}.

We find that the above $F_0$
coincides with the prepotential
computed from the Seiberg--Witten curve in the last section.
We verified it by computing the series expansion of $F_0$ in
$Q$ independently by each of the methods
and comparing the coefficients
up to order $Q^{10}$.
In doing this, the following identities
\eqb
\lefteqn{
\varth_{ab}(m+z,\tau)\varth_{ab}(m-z,\tau)
}\hspace{1.5em}\nn\\
\Eqn{=} \varth_{00}(2m,2\tau)\varth_{a0}(2z,2\tau)
 +(-1)^b\varth_{10}(2m,2\tau)\varth_{1-a,0}(2z,2\tau)\\[1ex]
\Eqn{=}\frac{1}{2}\left[
 \left(\frac{\varth_{00}(2m,2\tau)}{\varth_{a0}(0,2\tau)}
 +(-1)^b\frac{\varth_{10}(2m,2\tau)}{\varth_{1-a,0}(0,2\tau)}
 \right)\varth_{00}(z,\tau)^2\right.\nn\\
\label{thetaid2}
&&\hspace{1.5em}\left.{}+(-1)^a
 \left(\frac{\varth_{00}(2m,2\tau)}{\varth_{a0}(0,2\tau)}
 -(-1)^b\frac{\varth_{10}(2m,2\tau)}{\varth_{1-a,0}(0,2\tau)}
 \right)\varth_{01}(z,\tau)^2
\right]
\eqe
turn out to be useful. Using these identities one can
rewrite both the Seiberg--Witten curve and the Nekrasov-type expression
in such a way that all the dependence on $m_{cd}$ is expressed
through $\varth_{00}(2m_{cd},2\tau)$
and $\varth_{10}(2m_{cd},2\tau)$.
The comparison can then be made
in the same way as in the case of $\bfm=\bfzero$
by using the Taylor expansions of
the theta functions
\cite{Sakai:2012zq}.

As in \cite{Sakai:2012zq},
one can express $\calZ$ 
as a special case of
the
elliptic analogue of the Nekrasov partition function
for the $\grp{SU}(N)$ gauge theory
with $N_{\rm f}=2N$ fundamental hypermultiplets
\cite{Nekrasov:2002qd,Hollowood:2003cv}
\eqb
\lefteqn{
\ZSQCD{N}{2N}(\hbar;\varphi,\tau;a_1,\ldots,a_N;m_1,\ldots,m_{2N})
}\nn\\
\label{SQCDlike}
\Eqn{:=}\sum_{\bfR^{(N)}}
\left(-e^{2\pi i\varphi}\right)^{|\bfR^{(N)}|}
\prod_{k=1}^N
\prod_{(i,j)\in R_k}
\frac
{\prod_{n=1}^{2N}
 \varth_1\left(a_k+m_n+\tfrac{1}{2\pi}(j-i)\hbar,\tau\right)}
{\prod_{l=1}^N
\varth_1\left(a_k-a_l+\tfrac{1}{2\pi}h_{kl}(i,j)\hbar,\tau\right)^2}.
\eqe
In terms of this function,
(\ref{ZEstring}) 
can be expressed as
\[
\calZ = \ZSQCD{4}{8}
\left(\hbar;\varphi,\tau;0,\frac{1}{2},-\frac{1+\tau}{2},\frac{\tau}{2};
m_1,m_2,m_3,m_4,-m_1,-m_2,-m_3,-m_4\right).
\]
In the following sections we will see that
this type of notation provides us with an efficient, universal way
of expressing various Nekrasov-type formulas for specific cases.

Currently we do not have a good physical explanation
why the instanton counting of $\grp{SU}(4)\ N_{\rm f}=8$ type
accounts for the BPS spectrum of the E-string theory.
From the technical point of view
the elliptic analogue of
the Nekrasov partition function with four colors and eight flavors
is perfect
for reproducing the expansion $F_0 = \sum_{n=1}^\infty Z_nQ^n$
with $Z_1=\tfrac{1}{2}\eta^{-12}\sum_{k=1}^4\prod_{i=1}^8
\varth_k(\mu_i,\tau)$ \cite{Minahan:1998vr}.
No other known elliptic Nekrasov partition functions
\cite{Hollowood:2003cv}
seem to have an immediate connection with the above form of $Z_1$.
For particular values of Wilson line parameters, however,
one can express $\calZ$ in terms of
the elliptic analogues of the $\grp{SU}(N)\ N_{\rm f}=2N$
Nekrasov partition functions
with $N=3,2$, as we will see in the next subsection.
We have not examined whether
the BPS counting of the E-string theory has any connection with
the instanton counting of other types of gauge groups, for which 
no explicit elliptic Nekrasov partition functions are known.

The prepotential $F_0$ for the E-string theory represents
the genus zero topological string amplitude
for a family of local $\tfrac{1}{2}$K3 \cite{Minahan:1998vr}.
However, as was
observed in the case of $\bfm=\bfzero$ \cite{Sakai:2012zq},
higher order parts of the expansion
$\ln\calZ=\frac{1}{2}F_0\hbar^{-2}+\cdots$ do not give
higher genus amplitudes
\cite{Mohri:2001zz,Hosono:2002xj,Sakai:2011xg}.
The disagreement can be clearly seen as
the difference of modular anomalies.
With the help of (\ref{thetaid2})
one immediately sees that $\calZ$ exhibits the same
modular anomaly as in the case of $\bfm=\bfzero$.
This deviates from the modular anomaly of
the all-genus topological string partition function
for the local $\frac{1}{2}$K3
starting at genus one.

\subsection{Reductions to sums over fewer partitions}

For particular values of the Wilson line parameters
the above Nekrasov-type sum over partitions
reduces to that over fewer partitions.

Let us first consider the case where one of the four Wilson line
parameters is set to be zero,
\[
\bfm = (0,m_{10},m_{00},m_{01}).
\]
In this case, 
the product in the sum in (\ref{ZEstring}) vanishes
for any $\bfR^{(4)}$ with $R_{11}\ne\{0\}$.
This is because 
the Young diagram of $R_{11}\ne\{0\}$ always contains
a box at $(i,j)=(1,1)$,
where the theta functions in the numerator
become $\varth_{11}(0,\tau)=0$ for $(c,d)=(1,1)$.
Hence, $\calZ$ is actually a sum over three partitions
\[
\bfR^{(3)}=(R_{10},R_{00},R_{01}).
\]
This structure has already been found in the case of
no Wilson line parameters \cite{Sakai:2012zq}.
Furthermore, recall that for any function $f(x)$
the following identity holds:
\[
\prod_{(i,j)\in R_k}f\left(h_{k,l}(i,j)\right)
=\prod_{(i,j)\in R_k}f\left(j-i\right)
\quad\mbox{if}\quad R_l=\{0\}.
\]
This identity can be easily shown by regarding the product over
$j=1,\ldots,\nu_{k,i}$ as that over
$\tj:=\nu_{k,i}-j+1=1,\ldots,\nu_{k,i}$
on the left-hand side.
Due to this identity, one sees that
the expression for $\calZ$ reduces to the form
\[
\calZ=
\sum_{\bfR^{(3)}}
Q^{|\bfR^{(3)}|}
\prod_{(a,b),(c,d)}\,
\prod_{(i,j)\in R_{ab}}
\frac
{\varth_{ab}
 \left(\tfrac{1}{2\pi}(j-i)\hbar+m_{cd},\tau\right)
 \varth_{ab}
 \left(\tfrac{1}{2\pi}(j-i)\hbar-m_{cd},\tau\right)
}
{\varth_{1-|a-c|,1-|b-d|}
 \left(\tfrac{1}{2\pi}h_{ab,cd}(i,j)
       \hbar,\tau\right)^2}.
\]
This is almost identical to (\ref{ZEstring}),
except that the sum is now over $\bfR^{(3)}$
and indices $(a,b),(c,d)$ take values $(1,0),(0,0),(0,1)$ only.
In terms of the elliptic Nekrasov partition function (\ref{SQCDlike}),
the above simplification is expressed as
\eqb
\calZ
\Eqn{=}
\ZSQCD{4}{8}
\left(\hbar;\varphi,\tau;0,\frac{1}{2},-\frac{1+\tau}{2},\frac{\tau}{2};
0,m_{10},m_{00},m_{01},0,-m_{10},-m_{00},-m_{01}\right)\nn\\
\Eqn{=}
\ZSQCD{3}{6}
\left(\hbar;\varphi,\tau;\frac{1}{2},-\frac{1+\tau}{2},\frac{\tau}{2};
m_{10},m_{00},m_{01},-m_{10},-m_{00},-m_{01}\right).
\eqe
As we will see in the next section,
this simplified $\calZ$ encompasses all the cases of global symmetries
$E_n\oplus A_{8-n}$ with $n=8,7,6,5$ and $D_8$.

Next, let us further restrict ourselves to the cases with
\[\label{twofreems}
\bfm = \left(0,\frac{1}{2},m_1,m_2\right).
\]
In this setting, the expression for $\calZ$ reduces to the form
\[
\calZ=
\sum_{\bfR^{(2)}}
Q^{|\bfR^{(2)}|}
\prod_{k,l=1}^2\,
\prod_{(i,j)\in R_k}
\frac
{\varth_{k+2}\left(\tfrac{1}{2\pi}(j-i)\hbar+m_l,\tau\right)
 \varth_{k+2}\left(\tfrac{1}{2\pi}(j-i)\hbar-m_l,\tau\right)}
{\varth_{|k-l|+1}\left(\tfrac{1}{2\pi}h_{kl}(i,j)\hbar,\tau\right)^2},
\qquad
\]
where $\bfR^{(2)}=(R_1,R_2)$.
In terms of the elliptic Nekrasov partition function (\ref{SQCDlike}),
$\calZ$ with (\ref{twofreems}) can be expressed as
\eqb
\calZ
\Eqn{=}
\ZSQCD{4}{8}
\left(\hbar;\varphi,\tau;
0,\frac{1}{2},-\frac{1+\tau}{2},\frac{\tau}{2};
0,\frac{1}{2},m_1,m_2,0,-\frac{1}{2},-m_1,-m_2\right)\nn\\
\Eqn{=}
\ZSQCD{3}{6}
\left(\hbar;\varphi,\tau;
\frac{1}{2},-\frac{1+\tau}{2},\frac{\tau}{2};
\frac{1}{2},m_1,m_2,-\frac{1}{2},-m_1,-m_2\right)\nn\\
\Eqn{=}
\ZSQCD{2}{4}
\left(\hbar;\varphi,\tau;
-\frac{1+\tau}{2},\frac{\tau}{2};
m_1,m_2,-m_1,-m_2\right).
\eqe
As we will see in the next section,
the cases of global symmetries
$E_7\oplus A_1,\,E_5\oplus A_3$ and $D_8$
are realized as special cases of this setting.

Furthermore, if we set
\[
\bfm = \left(0,\frac{1}{2},-\frac{1+\tau}{2},\frac{\tau}{2}\right),
\]
$\calZ$ vanishes.
This is consistent with the fact that
the corresponding unbroken global symmetry
is $D_4\oplus D_4$ and the Seiberg--Witten curve in this case
describes a constant elliptic fibration
over the moduli space \cite{Eguchi:2002nx}.

\section{Two-parameter families}

In this section we consider the cases in which the Lie algebra
of the unbroken global symmetry is
$E_{9-N}\oplus A_{N-1}$ with $N=1,2,3,4$ or $D_8$.
These cases are of particular interest
because the prepotential in each case generates
Gromov--Witten invariants
associated with the $E_{9-N}$ del Pezzo surface
or $\bbP^1\times \bbP^1$ embedded in a Calabi--Yau threefold.
In particular, we consider 
two-parameter families of Calabi--Yau,
whose prepotentials depend
not only on K\"ahler modulus $\varphi$
but also on another K\"ahler modulus $\tau$.
These two-parameter families have been studied
by means of mirror symmetry
\cite{Klemm:1996hh,Mohri:2001zz}.

All of the above global symmetries
are maximal regular subalgebras of $E_8$
and one can easily find the corresponding values
of Wilson line parameters \cite{Eguchi:2002nx}.
As we saw in section~2,
different values of Wilson line parameters $\bfm$
related by the transformations (\ref{minterchange})--(\ref{mtransl})
correspond to the same Seiberg--Witten curve.
We will present a representative of $\bfm$
for each of these cases.

Substituting each of these $\bfm$ into our general expression
given by (\ref{N_f=4curve})--(\ref{Mthetarel})
one obtains the Seiberg--Witten curve for each of the cases.
The result can be simplified
by making use of theta function identities.
We will present the final form of the curve
after making a shift of variables $x$ and $u$.
The shift leads to a further simplification.
The curve without the shift can easily be recovered
by first eliminating the quadratic term in $x$ by a shift of $x$
and then eliminating the cubic part in $u$ of the linear term in $x$
by a shift of $u$.
The Seiberg--Witten curve for the E-string theory describes
an elliptic fibration over $\bbP^1$ with singular fibers.
Using the Weierstrass form of the curve
one can easily check that the types of singular fibers
correspond precisely to the simple Lie algebras constituting
the unbroken global symmetry \cite{Eguchi:2002nx}.

We will also present explicit
Nekrasov-type expressions for each of the cases.
The prepotential is obtained from $\calZ$ through (\ref{F0Estring}).
Following \cite{Minahan:1998vr}
we introduce the winding number expansion of the prepotential by
\[\label{windingexp}
F_0(\varphi,\tau)=\sum_{n=1}^\infty Q^n Z_n(\tau).
\]
$Z_n$ for $E_{9-N}\oplus A_{N-1}$ with $N=1,2,3,4$
can be expressed in terms of $E_2(\tau)$ and
modular forms of
$\Gamma_1(N)=\Bigl\{
\Bigl(\begin{array}{cc}a&b\\ c&d\end{array}\Bigr)
\in\grp{SL}(2,\bbZ)\Bigm| a\equiv d\equiv 1,\,c\equiv 0 \mod N
\Bigr\}$.
We will introduce generators $\gA{N},\gB{N}$
of modular forms of $\Gamma_1(N)$
as well as a function $\gL{N}$
and present explicit forms of $Z_n$ for small $n$.
In the $D_8$ case, $Z_n$ are expressed in terms
of $E_2(\tau)$ and modular forms of $\Gamma_1(2)$. 

In each of the above cases, the prepotential
can be expressed as
\[
F_0(\varphi,\tau)=
\sum_{n=1}^\infty\sum_{k=0}^\infty N_{n,k}
\sum_{m=1}^\infty\frac{1}{m^3}e^{2\pi im(n\varphi+k\tau)}.
\]
Integer $N_{n,k}$ represents the multiplicity of
BPS E-strings wound around one of the circles of the toroidal
compactification with winding number $n$ and momentum $k$.
Up to an overall normalization
the values of $N_{n,n}$
turn out to be equal to
the genus zero Gromov--Witten invariants
associated with
the $E_{9-N}$ del Pezzo surface or
$\bbP^1\times \bbP^1$ \linebreak[4]
embedded in a Calabi--Yau threefold
\cite{Klemm:1996hh,Lerche:1996ni,Chiang:1999tz,Katz:1999xq}.
Our formulas also generate invariants $N_{n,k}$ with $k\ne n$,
where $k$ is the degree associated with
the homology class of the elliptic fiber.
It would be very interesting to see how
our combinatorial expressions are related to
the geometric computation of these invariants \cite{Katz:1999xq}.

\subsection{$E_8$}

Let us first consider the case with an $E_8$ global symmetry.
This is the case originally discussed in \cite{Sakai:2012zq}
and is realized by the trivial Wilson line parameters
\[
\bfm=\left(0,0,0,0\right).
\]
The corresponding Seiberg--Witten curve is given by
\[
y^2=4x^3-\frac{1}{12}E_4u^4x-\frac{1}{216}E_6u^6+4u^5.
\]
The Nekrasov-type expression can be written as
\eqb
\calZ
\Eqn{=}
\ZSQCD{3}{6}
\left(\hbar;\varphi,\tau;\frac{1}{2},-\frac{1+\tau}{2},\frac{\tau}{2};
0,0,0,0,0,0\right)\\
\Eqn{=}
\sum_{\bfR^{(3)}}
Q^{|\bfR^{(3)}|}
\prod_{(a,b),(c,d)}\,
\prod_{(i,j)\in R_{ab}}
\frac
{\varth_{ab}
 \left(\tfrac{1}{2\pi}(j-i)\hbar,\tau\right)^2}
{\varth_{1-|a-c|,1-|b-d|}
 \left(\tfrac{1}{2\pi}h_{ab,cd}(i,j)
       \hbar,\tau\right)^2}.
\eqe
Here the set of indices $(a,b),(c,d)$
take values $(1,0),(0,0),(0,1)$ and
we let the three partitions be denoted by
$\bfR^{(3)}=(R_{10},R_{00},R_{01})$.
The first three coefficients of the expansion (\ref{windingexp})
are
\eqb
Z_1\Eqn{=}\gL{1}\gA{1},\nn\\
Z_2\Eqn{=}\gL{1}^2\gA{1}\left(\frac{\gA{1}E_2+2\gB{1}}{24}\right),\nn\\
Z_3\Eqn{=}\gL{1}^3\gA{1}\left(\frac{
  54\gA{1}^2E_2^2+216\gA{1}\gB{1}E_2+109\gA{1}^3+197\gB{1}^2}{15552}
  \right),
\eqe
where
\[
\gA{1}:=E_4,\qquad \gB{1}:=E_6
\]
and
\[
\gL{1}:=\frac{1}{\eta^{12}}.
\]
These $Z_n$ agree with the original results \cite{Minahan:1997ct}.
Table \ref{NnkforE8} shows the values of $N_{n,k}$
for low $n$ and $k$. These numbers were originally computed
by using mirror symmetry \cite{Klemm:1996hh}.

\begin{table}[t]
\[
\begin{array}{|c|rrrrrrrr|}\hline
&k&0&1&2&3&4&5&\cdots\\ \hline
n&&&&&&&&\\
1&&1&252&5130&54760&419895&2587788&\\
2&&0&0&-9252&-673760&-20534040&-389320128&\\
3&&0&0&0&848628&115243155&6499779552&\\
4&&0&0&0&0&-114265008&-23064530112&\\
5&&0&0&0&0&0&18958064400&\\
\vdots &&&&&&&&\ddots\\ \hline
\end{array}
\nn
\]
\caption{BPS multiplicities $N_{n,k}$
for the $E_8$ case.
\label{NnkforE8}}
\end{table}
%

\subsection{$E_7\oplus A_1$}

The $E_7\oplus A_1$ symmetry is realized by
the following Wilson line parameters
\[\label{E7A1masses}
\bfm=\left(0,0,0,\frac{1}{2}\right).
\]
The Seiberg--Witten curve is given by
\[
y^2=4x^3+\left(\varth_3^4+\varth_4^4\right)u^2x^2
+\left(\frac{\varth_3^4\varth_4^4}{4}u
  -\frac{16}{\varth_3^2\varth_4^2}\right)u^3x.
\]
The Nekrasov-type expression in this case takes
a remarkably simple form
\eqb
\calZ
\Eqn{=}
\ZSQCD{2}{4}
\left(\hbar;\varphi,\tau;-\frac{1+\tau}{2},\frac{\tau}{2};
0,0,0,0\right)\\
\Eqn{=}
\sum_{\bfR^{(2)}}
Q^{|\bfR^{(2)}|}
\prod_{k,l=1}^2\,
\prod_{(i,j)\in R_k}
\frac
{\varth_{k+2}\left(\tfrac{1}{2\pi}(j-i)\hbar,\tau\right)^2}
{\varth_{|k-l|+1}\left(\tfrac{1}{2\pi}h_{kl}(i,j)\hbar,\tau\right)^2}.
\eqe
The first three coefficients of the expansion (\ref{windingexp})
are
\eqb
Z_1\Eqn{=}\gL{2}\gA{2},\nn\\
Z_2\Eqn{=}\gL{2}^2\gA{2}
  \left(\frac{\gA{2}E_2-\gA{2}^2+3\gB{2}}{24}\right),\nn\\
Z_3\Eqn{=}\gL{2}^3\gA{2}
  \left(\frac{6\gA{2}^2E_2^2-12\gA{2}^3E_2+36\gA{2}\gB{2}E_2
  +16\gA{2}^4-33\gA{2}^2\gB{2}+51\gB{2}^2}{1728}\right),\quad
\eqe
where
\[\label{E7gen}
\gA{2} := 
\frac{1}{2}\left(\varth_3^4+\varth_4^4\right),\qquad
\gB{2} :=\varth_3^4\varth_4^4=\frac{\eta(\tau)^{16}}{\eta(2\tau)^8}
\]
and
\[
\gL{2} := \frac{\varth_3^2\varth_4^2}{\eta^{12}}
=\frac{1}{\eta(\tau)^{4}\eta(2\tau)^{4}}.
\]
As expected, these $Z_n$ are in agreement with
the results obtained in \cite{Mohri:2001zz}.
Table \ref{NnkforE7} shows the values of $N_{n,k}$
for low $n$ and $k$.
The values of $N_{n,k}$ multiplied by two agree with
the rational instanton numbers of the $E_7$ model
in \cite{Mohri:2001zz}.

\begin{table}[t]
\[
\begin{array}{|c|rrrrrrrr|}\hline
&k&0&1&2&3&4&5&\cdots\\ \hline
n&&&&&&&&\\
1&&1&28&138&680&2359&7980&\\
2&&0&0&-136&-2272&-23208&-167872&\\
3&&0&0&0&1620&50067&824544&\\
4&&0&0&0&0&-29216&-1316544&\\
5&&0&0&0&0&0&651920&\\
\vdots &&&&&&&&\ddots\\ \hline
\end{array}
\nn
\]
\caption{BPS multiplicities $N_{n,k}$
for the $E_7 \oplus A_1$ case.
\label{NnkforE7}}
\end{table}
%

\subsection{$E_6\oplus A_2$}

The $E_6\oplus A_2$ symmetry is realized by
the following Wilson line parameters
\[
\bfm=\left(0,\frac{1}{3},\frac{1}{3},\frac{1}{3}\right).
\]
The Seiberg--Witten curve is given by
\[
y^2=
4x^3+3\gA{3}^2u^2x^2
+\frac{2}{3}\gA{3}\left(\gB{3}u-\frac{27}{\gB{3}}\right)u^3x
+\frac{1}{27}\left(\gB{3}u-\frac{27}{\gB{3}}\right)^2u^4,
\]
where
\[\label{E6gen}
\gA{3} := \varth_3(0,2\tau)\varth_3(0,6\tau)
            +\varth_2(0,2\tau)\varth_2(0,6\tau),\qquad
\gB{3} := \frac{\eta(\tau)^9}{\eta(3\tau)^3}.
\]
The Nekrasov-type expression is given by
\eqb
\calZ
\Eqn{=}
\ZSQCD{3}{6}
\left(\hbar;\varphi,\tau;\frac{1}{2},-\frac{1+\tau}{2},\frac{\tau}{2};
\frac{1}{3},\frac{1}{3},\frac{1}{3},
-\frac{1}{3},-\frac{1}{3},-\frac{1}{3}\right)\\
\Eqn{=}
\sum_{\bfR^{(3)}}
Q^{|\bfR^{(3)}|}
\prod_{(a,b),(c,d)}\,
\prod_{(i,j)\in R_{ab}}
\frac
{\varth_{ab}
 \left(\tfrac{1}{2\pi}(j-i)\hbar+\tfrac{1}{3},\tau\right)
 \varth_{ab}
 \left(\tfrac{1}{2\pi}(j-i)\hbar-\tfrac{1}{3},\tau\right)
}
{\varth_{1-|a-c|,1-|b-d|}
 \left(\tfrac{1}{2\pi}h_{ab,cd}(i,j)
       \hbar,\tau\right)^2},\qquad\ 
\eqe
where $(a,b),(c,d)=(1,0),(0,0),(0,1)$ and
$\bfR^{(3)}=(R_{10},R_{00},R_{01})$.
The first three coefficients of the expansion (\ref{windingexp})
are
\eqb
Z_1\Eqn{=}\gL{3}\gA{3},\nn\\
Z_2\Eqn{=}\gL{3}^2\gA{3}\left(\frac{\gA{3}E_2+2\gB{3}}{24}\right),\nn\\
Z_3\Eqn{=}\gL{3}^3\gA{3}\left(\frac{
 27\gA{3}^2E_2^2+108\gA{3}\gB{3}E_2+45\gA{3}^6
 -4\gA{3}^3\gB{3}+112\gB{3}^2}{7776}\right),
\eqe
where
\[
\gL{3} := \frac{\gB{3}}{\eta^{12}}
=\frac{1}{\eta(\tau)^3\eta(3\tau)^3}.
\]
Table \ref{NnkforE6} shows the values of $N_{n,k}$
for low $n$ and $k$.
The values of $N_{n,k}$ multiplied by three agree with
the rational instanton numbers of the $E_6$ model
in \cite{Mohri:2001zz}.

\begin{table}[t]
\[
\begin{array}{|c|rrrrrrrr|}\hline
&k&0&1&2&3&4&5&\cdots\\ \hline
n&&&&&&&&\\
1&&1&9&27&85&234&567&\\
2&&0&0&-18&-164&-1026&-4968&\\
3&&0&0&0&81&1377&13365&\\
4&&0&0&0&0&-576&-14040&\\
5&&0&0&0&0&0&5085&\\
\vdots &&&&&&&&\ddots\\ \hline
\end{array}
\nn
\]
\caption{BPS multiplicities $N_{n,k}$
for the $E_6 \oplus A_2$ case.
\label{NnkforE6}}
\end{table}
%

\subsection{$E_5\oplus A_3$}

The $E_5\oplus A_3$ symmetry is realized by
the following Wilson line parameters
\[
\bfm=\left(0,\frac{1}{4},\frac{1}{4},\frac{1}{2}\right).
\]
The Seiberg--Witten curve and the Nekrasov-type expression
in this case are given respectively by
\[
y^2=4x^3
+\left(\left(\varth_3^4+\varth_4^4\right)u
  +\frac{64}{(\varth_3^2+\varth_4^2)\varth_3^3\varth_4^3}\right)ux^2
+\left(\frac{\varth_3^2\varth_4^2}{2}u
  -\frac{16}{(\varth_3^2+\varth_4^2)\varth_3^3\varth_4^3}\right)^2u^2x
\]
and
\eqb
\calZ
\Eqn{=}
\ZSQCD{2}{4}
\left(\hbar;\varphi,\tau;-\frac{1+\tau}{2},\frac{\tau}{2};
\frac{1}{4},\frac{1}{4},-\frac{1}{4},-\frac{1}{4}\right)\\
\Eqn{=}
\sum_{\bfR^{(2)}}
Q^{|\bfR^{(2)}|}
\prod_{k,l=1}^2\,
\prod_{(i,j)\in R_k}
\frac
{\varth_{k+2}\left(\tfrac{1}{2\pi}(j-i)\hbar+\tfrac{1}{4},\tau\right)
 \varth_{k+2}\left(\tfrac{1}{2\pi}(j-i)\hbar-\tfrac{1}{4},\tau\right)}
{\varth_{|k-l|+1}\left(\tfrac{1}{2\pi}h_{kl}(i,j)\hbar,\tau\right)^2}.
\qquad
\eqe
The first three coefficients of the expansion (\ref{windingexp})
are
\eqb
Z_1\Eqn{=}\gL{4},\nn\\
Z_2\Eqn{=}\gL{4}^2\left(\frac{E_2+\gA{4}+\gB{4}}{24}\right),\nn\\
Z_3\Eqn{=}\gL{4}^3\left(\frac{
  3E_2^2+6\gA{4}E_2+6\gB{4}E_2+8\gA{4}^2+4\gA{4}\gB{4}+5\gB{4}^2
  }{864}\right),
\eqe
where
\[
\gA{4}:=\varth_3(0,2\tau)^4
=\frac{\eta(2\tau)^{20}}{\eta(\tau)^8\eta(4\tau)^8},\qquad
\gB{4}:= \varth_4(0,2\tau)^4
=\frac{\eta(\tau)^8}{\eta(2\tau)^4}
\]
and
\[
\gL{4} := \frac{\varth_3(0,2\tau)^2\varth_4(0,2\tau)^6}{\eta^{12}}
=\frac{(\varth_3^2+\varth_4^2)\varth_3^3\varth_4^3}{2\eta^{12}}
=\frac{\eta(2\tau)^4}{\eta(\tau)^4\eta(4\tau)^4}.
\]
Table \ref{NnkforE5} shows the values of $N_{n,k}$
for low $n$ and $k$.
The values of $N_{n,k}$ multiplied by four agree with
the rational instanton numbers of the $E_5$ model
in \cite{Mohri:2001zz}.

\begin{table}[t]
\[
\begin{array}{|c|rrrrrrrr|}\hline
&k&0&1&2&3&4&5&\cdots\\ \hline
n&&&&&&&&\\
1&&1&4&10&24&55&116&\\
2&&0&0&-5&-32&-152&-576&\\
3&&0&0&0&12&147&1056&\\
4&&0&0&0&0&-48&-832&\\
5&&0&0&0&0&0&240&\\
\vdots &&&&&&&&\ddots\\ \hline
\end{array}
\nn
\]
\caption{BPS multiplicities $N_{n,k}$
for the $E_5 \oplus A_3$ case.
\label{NnkforE5}}
\end{table}
%

\subsection{$D_8$}

The $D_8$ symmetry is realized by
the following Wilson line parameters
\[
\bfm=\left(0,0,\frac{1}{2},\frac{1}{2}\right).
\]
The Seiberg--Witten curve and the Nekrasov-type expression
in this case are given respectively by
\[
y^2=4x^3+\left(\left(\varth_3^4+\varth_4^4\right)u
              +\frac{64}{\varth_3^4\varth_4^4}\right)ux^2
+\frac{\varth_3^4\varth_4^4}{4}u^4x
\]
and
\eqb
\calZ
\Eqn{=}
\ZSQCD{2}{4}
\left(\hbar;\varphi,\tau;-\frac{1+\tau}{2},\frac{\tau}{2};
0,0,\frac{1}{2},-\frac{1}{2}\right)\\
\Eqn{=}
\sum_{\bfR^{(2)}}
Q^{|\bfR^{(2)}|}
\prod_{k,l=1}^2\,
\prod_{(i,j)\in R_k}
\frac
{\varth_3\left(\tfrac{1}{2\pi}(j-i)\hbar,\tau\right)
 \varth_4\left(\tfrac{1}{2\pi}(j-i)\hbar,\tau\right)}
{\varth_{|k-l|+1}\left(\tfrac{1}{2\pi}h_{kl}(i,j)\hbar,\tau\right)^2}.
\qquad
\eqe
The first three coefficients of the expansion (\ref{windingexp})
are
\eqb
Z_1\Eqn{=}\tgL{2},\nn\\
Z_2\Eqn{=}\tgL{2}^2\left(\frac{E_2+2\gA{2}}{24}\right),\nn\\
Z_3\Eqn{=}\tgL{2}^3\left(\frac{
  6E_2^2+24\gA{2}E_2+25\gA{2}^2+9\gB{2}}{1728}\right),
\eqe
where
$\gA{2},\gB{2}$ are defined in (\ref{E7gen}) and
\[
\tgL{2} := \frac{\varth_3^4\varth_4^4}{\eta^{12}}
=\frac{\eta(\tau)^4}{\eta(2\tau)^8}
=\frac{2^4}{\varth_2^4}.
\]
Table \ref{NnkforD8} shows the values of $N_{n,k}$
for low $n$ and $k$.
We observe that the values of $N_{n,n}$ are related to
the genus zero Gromov--Witten invariants
of the local $\bbP^1\times\bbP^1$ as
\[
N_{n,n}=\sum_{n_1+n_2=n}N^{\bbP^1\times\bbP^1}_{n_1,n_2}.
\]
Note that the values of $N_{n,k}$ for the $D_8$ case
are very similar to those for the $E_5\oplus A_3$ case.
This has been explained by the similarity
between the Picard--Fuchs operators
for $\bbP^1\times \bbP^1$
(which is equal to the quadric surface in $\bbP^3$)
and those for the $E_5$ del Pezzo surface \cite{Lerche:1996ni}.
From the point of view of the E-string theory
this may be explained by
the similarity between the Weyl orbits of
$D_8$ and those of
$E_5\oplus A_3\cong D_5\oplus D_3$.

\begin{table}[t]
\[
\begin{array}{|c|rrrrrrrr|}\hline
&k&0&1&2&3&4&5&\cdots\\ \hline
n&&&&&&&&\\
1&&1&-4&10&-24&55&-116&\\
2&&0&0&-4&32&-152&576&\\
3&&0&0&0&-12&147&-1056&\\
4&&0&0&0&0&-48&832&\\
5&&0&0&0&0&0&-240&\\
\vdots &&&&&&&&\ddots\\ \hline
\end{array}
\nn
\]
\caption{BPS multiplicities $N_{n,k}$
for the $D_8$ case.
\label{NnkforD8}}
\end{table}
%

\vspace{3ex}

\begin{center}
  {\bf Acknowledgments}
\end{center}

The author would like to thank T.~Eguchi, K.~Hosomichi,
A.~Prudenziati, \linebreak[4]
Y.~Tachikawa and S.~Terashima
for discussions.
He is grateful to the Institut de Physique Th\'eorique
at CEA Saclay for hospitality.
The author is the Yukawa Fellow and his work is supported
in part by Yukawa Memorial Foundation.
His work is also
supported in part by Grant-in-Aid
for Scientific Research from the Japan Ministry of Education, Culture, 
Sports, Science and Technology.

\vspace{3ex}

\newpage

\appendix

\section{Conventions of special functions}

The Jacobi theta functions are defined as
\[
\varth_{ab}(z,\tau)
 :=\sum_{n\in \bbZ}\exp\left[
  \pi i\left(n+\frac{a}{2}\right)^2\tau
 +2\pi i\left(n+\frac{a}{2}\right)\left(z+\frac{b}{2}\right)\right],
\]
where $a,b$ take values $0,1$. We also use the notation
\begin{align}
\hspace{3em}
\varth_1(z,\tau) &:= -\varth_{11}(z,\tau),&
\varth_2(z,\tau) &:= \varth_{10}(z,\tau),\hspace{5em}\nn\\
\varth_3(z,\tau) &:= \varth_{00}(z,\tau),&
\varth_4(z,\tau) &:= \varth_{01}(z,\tau).
\end{align}
The Dedekind eta function is defined as
\[
\eta(\tau) := q^{1/24}\prod_{n=1}^\infty (1-q^n),
\]
where $q:=e^{2\pi i\tau}$.
The Eisenstein series are given by
\[
E_{2n}(\tau)
 =1+\frac{2}{\zeta(1-2n)}
 \sum_{k=1}^{\infty}\frac{k^{2n-1}q^k}{1-q^k}.
\]
We often abbreviate $\varth_k(0,\tau),\,\eta(\tau),\,E_{2n}(\tau)$
as $\varth_k,\,\eta,\,E_{2n}$ respectively.

\vspace{3ex}


\renewcommand{\section}{\subsection}
\renewcommand{\refname}

\end{document}